\documentclass[cits]{PoS}
\usepackage{wasysym}

\title{Lowering the low-energy threshold of xenon detectors}

\ShortTitle{Lowering the low-energy threshold of xenon detectors}

\author{\speaker{P.~Sorensen},$^a$ J.~Angle,$^{b,c}$ E.~Aprile,$^d$ F.~Arneodo,$^e$ L.~Baudis,$^{b,c}$ A.~Bernstein,$^a$ A.~Bolozdynya,$^f$ L.C.C.~Coelho,$^g$ C.E.~Dahl,$^{f,h}$ L.~DeViveiros,$^i$ A.D.~Ferella,$^{c,e}$ L.M.P.~Fernandes,$^g$ S.~Fiorucci,$^i$ R.J.~Gaitskell,$^i$ K.L.~Giboni,$^d$ R.~Gomez,$^j$ R.~Hasty,$^k$ L.~Kastens,$^k$ J.~Kwong,$^{f,h}$ J.A.M.~Lopes,$^g$ N.~Madden,$^a$ A.~Manalaysay,$^{b,c}$ A.~Manzur,$^k$ D.N.~McKinsey,$^k$ M.E.~Monzani,$^d$ K.~Ni,$^k$ U.~Oberlack,$^j$ J.~Orboeck,$^l$ G.~Plante,$^d$ R.~Santorelli,$^d$ J.M.F.~dos~Santos,$^g$ S.~Schulte,$^l$ P.~Shagin,$^j$ T.~Shutt,$^f$ C.~Winant,$^a$ M.~Yamashita$^d$\\
        \llap{$^a$}Lawrence Livermore National Laboratory, 7000 East Ave., Livermore, CA 94550, USA\\
        \llap{$^b$}Department of Physics, University of Florida, Gainesville, FL 32611, USA\\
        \llap{$^c$}Physics Institute, University of Z\"urich, Winterthurerstrasse 190, CH-8057,  Z\"urich, Switzerland\\
        \llap{$^d$}Department of Physics, Columbia University, New York, NY 10027, USA\\
        \llap{$^e$}Gran Sasso National Laboratory, Assergi, L'Aquila, 67010, Italy\\
        \llap{$^f$}Department of Physics, Case Western Reserve University, Cleveland, OH 44106, USA\\
        \llap{$^g$}Department of Physics, University of Coimbra, R. Larga, 3004-516, Coimbra, Portugal\\
        \llap{$^h$}Department of Physics, Princeton University, Princeton, NJ 08540, USA\\
        \llap{$^i$}Department of Physics, Brown University, Providence, RI 02912, USA\\
        \llap{$^j$}Department of Physics and Astronomy, Rice University, Houston, TX 77251, USA\\
        \llap{$^k$}Department of Physics, Yale University, New Haven, CT 06511, USA\\
        \llap{$^l$}Department of Physics, RWTH Aachen University, Aachen, 52074, Germany\\
        E-mail: \email{pfs@llnl.gov}}

\abstract{We show that the energy threshold for nuclear recoils in the XENON10 dark matter search data can be lowered to $\sim1$~keV, by using only the ionization signal.  In other words, we make no requirement that a valid event contain a primary scintillation signal.  We therefore relinquish incident particle type discrimination, which is based on the ratio of ionization to scintillation in liquid xenon.  This method compromises the detector's ability to precisely determine the $z$ coordinate of a particle interaction.  However, we show for the first time that it is possible to discriminate bulk events from surface events based solely on the ionization signal.}

\FullConference{Identification of Dark Matter 2010-IDM2010\\
		July 26-30, 2010\\
		Montpellier France}

\begin{document}
\section{Introduction} \label{intro}
The XENON10 detector \cite{2010aprile} is a liquid xenon time-projection chamber with a an active target mass of 13.7~kg (15~cm height and 10~cm radius).  It was designed to directly detect galactic dark matter particles which scatter off xenon nuclei.  Typical velocities of halo-bound dark matter particles are of order 10$^{-3}$c.  This leads to the prediction of featureless exponential recoil energy spectra for spin-independent elastic scattering of dark matter particles on a xenon target \cite{1996jungman}.  Typical nuclear recoil energies are few keV.  A particle interaction in liquid xenon results in a prompt scintillation signal (S1) and an ionization signal (S2).  The generation and collection of these signals are discussed in detail in \cite{2010aprile}.  A key feature of the XENON10 detector is the event-by-event discrimination between incident particle type, based on the ratio S2/S1.

The energy threshold of previously reported XENON10 data  \cite{2009angle,2008angle,2008angle2} depends on the primary scintillation efficiency of liquid xenon for nuclear recoils $\mathcal{L}_{eff}$ \cite{2010manzur,2009aprile}, and for a conservative assumption of the energy dependence of $\mathcal{L}_{eff}$ \cite{2010manzur}, is about 5~keV.  This energy threshold is dictated by the collection of primary scintillation photons following a nuclear recoil in the xenon target.  But it is possible to obtain a lower energy threshold from the existing XENON10 data, using only the ionization signal.  We discuss an analysis of XENON10 dark matter search data, with the energy scale set by the detected ionization signal and no requirement that valid events contain both an S1 and an S2 signal.  This requires a compromise on two important aspects of the detector performance:  the ability to precisely reconstruct the $z$ coordinate of a particle interaction (which is normally obtained by the time delay between S1 and S2 signals), and the discrimination between incident particle types.

Calibration of the ionization energy scale for nuclear recoils in liquid xenon is a pre-requisite.  Direct measurements exist in the literature \cite{2010manzur,2006aprile} (and are shown in Fig. \ref{fig2}b), but do not extend to the smallest signals observed in our detector.  We describe a new measurement of the ionization yield of liquid xenon for nuclear recoils, down to a nuclear recoil energy of about 1~keVr.  We then describe a method for making an approximate determination of the $z$ coordinate of the interaction.  The method exploits the measured width of the S2 signal, which is broadened by electron diffusion as a cloud of electrons drifts across the liquid xenon target.  The precision is low ($\sim$cm), but is sufficient to discriminate bulk events from surface events.  This is important because the latter are more likely to result from radioactive background.  

\section{Ionization yield of liquid xenon for nuclear recoils} \label{qysec}
\begin{figure}
\centering
\includegraphics[width=0.99\textwidth]{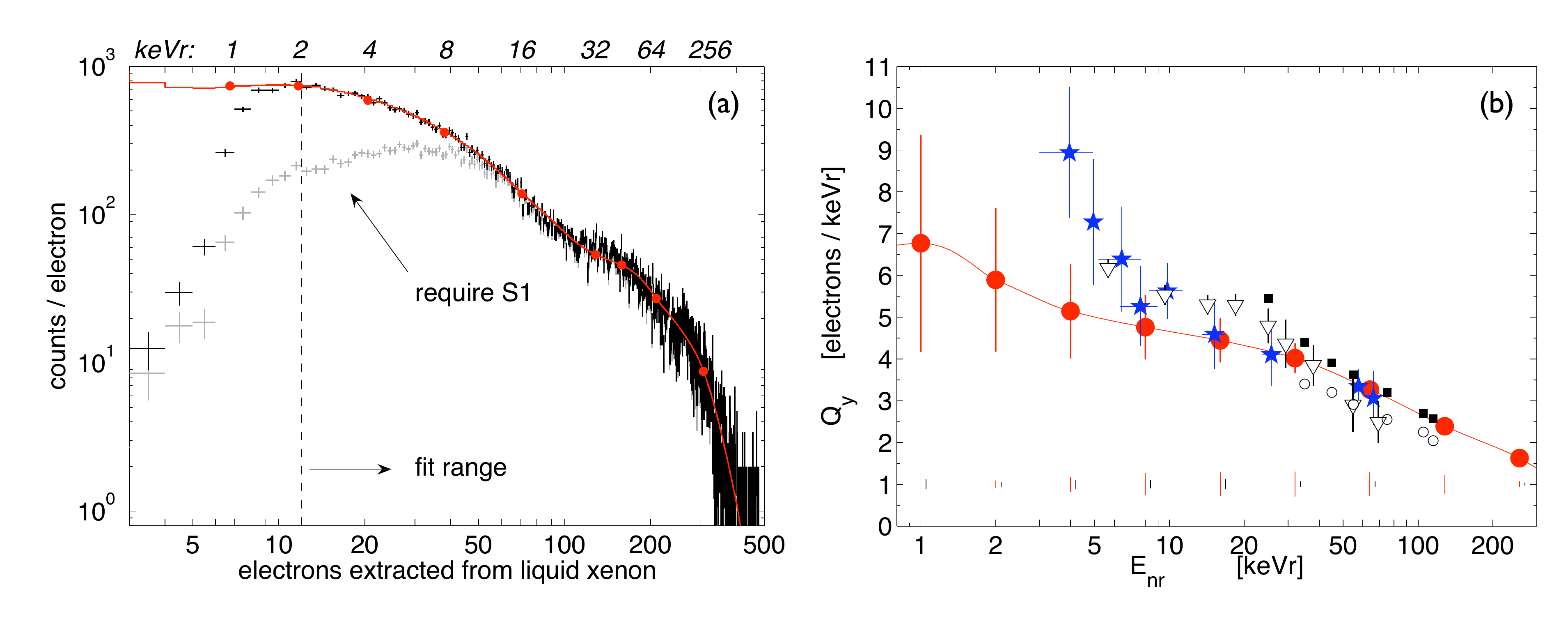}
\caption{{\bf (a)} Spectrum of the number of electrons extracted from single scatter elastic nuclear recoil interactions in the fiducial target defined by $r<8$~cm, with $1\sigma$ statistical uncertainty.  The best-fit Monte Carlo is shown as a continuous stair-step.  The energy values of the 9 spline points are identical in both plots.  Shown in light gray is the spectrum of events in which an S1 signal was found in the $80~\mu$s before the S2.  Approximate keVr-equivalent is indicated along the top. {\bf (b)} The ionization yield from nuclear recoils $\mathcal{Q}_y$ from the present work ($\newmoon$), with $1\sigma$ statistical uncertainty.  Systematic uncertainty is shown along the axis. Also shown are data from \cite{2010manzur} ($\bigstar$), \cite{2009sorensen} ($\triangledown$) and \cite{2006aprile} ($\fullmoon$ and $\blacksquare$ ).}
\label{fig1}
\end{figure}

The ionization yield $\mathcal{Q}_y$ was obtained directly from {\it in-situ} neutron calibration data.  The neutron calibration experiment and general analysis technique are both described in a previous study by the XENON10 collaboration \cite{2009sorensen}.  Note that here we circumvent the {\it false single scatter} pathology described in \cite{2009angle}, since our calibration technique relies only on the S2 signal.  The S2 spectrum for single elastic neutron scatters is shown in Fig. \ref{fig1}a, with 1$\sigma$ uncertainty.  The S2 signal was scaled to an absolute number of electrons via the measured 24~photoelectrons per liquid electron \cite{2010aprile}.  Only the most basic data quality cuts were applied, along with a fiducial cut $r<8$~cm.  No $z$ fiducial cut was applied because a substantial fraction of events at low energy have no S1 (and hence have indeterminate $z$).  The fiducial target volume used in this calibration thus differs from previous analyses \cite{2008angle,2008angle2,2009angle,2009sorensen} in that it considers the full $\Delta z=15$~cm active target (8.6~kg target mass), rather than just the central $\Delta z=9.3$~cm (5.4~kg target mass).  

With no self-shielding liquid xenon above or below the chosen xenon target volume, it is important to take account of possible gamma and beta background contamination in the nuclear recoil data. This background is assessed by counting the number of events within $\pm2\sigma$ of the electron recoil centroid, in the region around $30-35$~keVee.  The chosen energy window is above the tail of the elastic nuclear recoil distribution, and below the 40~keVee inelastic scatters \cite{2009sorensen}.  The electromagnetic background spectrum below about 50~keVee is predicted and observed to be flat with energy.  Therefore, we were able to subtract the resulting background prediction of about 0.75 counts/electron.  Considering the number of nuclear recoil events in the data sample shown in Fig. \ref{fig1}a, this is a very small correction.

In comparing the measured energy spectrum with the Monte Carlo predicted spectrum, the energy resolution in the electron signal was assumed to be Poisson in the number of detected electrons;  account was also taken of the 20\% 1$\sigma$ width of the single electron distribution \cite{2009dahl}.  The S2-sensitive trigger threshold for the neutron calibration data had full efficiency for events with at least 182~photoelectrons \cite{2010aprile}, or 7.6~S2~electrons.  This is valid for $r\leq3$~cm, and there is a slight radial dependence to the trigger such that by $r=8$~cm  the efficiency is unity for events with greater than 11~S2~electrons.  As a result, events with S2~$<12$~electrons were not used in the calibration analysis.  No attempt was made to model the S2 trigger efficiency, as is apparent in Fig. \ref{fig1}a from the discrepancy between data and simulation below $\sim8$ electrons.

The ionization yield $\mathcal{Q}_y$ was modeled as a continuous cubic Hermite spline interpolation as in \cite{2009sorensen}, and a maximum likelihood comparison was applied to find $\mathcal{Q}_y(\mbox{E}_{nr})$ as shown in Fig. \ref{fig1}b ($\newmoon$, with $1\sigma$ statistical uncertainty).  No constraints were applied to the spline other than the requirement that $\mathcal{Q}_y=0$ at E$_{nr}=0.1$~keVr.  The exact minimum recoil energy that can result in a single detectable electron is not known, but is limited by the ionization potential of xenon and by nuclear recoil quenching \cite{1963lindhard}.  Our chosen boundary condition is conservative, and our results are insensitive to a $\times5$ shift in this value.  The spline points were chosen at fixed values of keV nuclear recoil energy (keVr), and the results do not depend on the location of the spline points (as long as there are $\sim10$ points spanning the full energy range).  The best fit shown in Fig. \ref{fig1}a has $\chi^2/d.o.f.=414/425$.  For reference, the ionization yield of 122~keV gammas from a $^{57}$Co source were found to have an ionization yield of $46.5\pm7.0~(stat.)$~electrons/keV, for events with radial positions in the range $8.5<r<9.0$~cm.

Two sources of systematic uncertainty are indicated as vertical bars for each spline point, along the $x$-axis.  These arise from an assumed $\pm10\%$ uncertainty in the single electron calibration (left bar), and from uncertainty below 2~MeV in the spectrum of initial neutron energies E$_n$ from the AmBe neutron source \cite{1995marsh} (right bar).  The neutron energy spectrum was taken from Fig. 5 of \cite{1995marsh}, and the uncertainty was parameterized as $1\pm \mbox{exp}(-\mbox{E}_n-\frac{1}{2})$ for E$_n<2$~MeV.  The effect of this conservative assumption on the Monte Carlo nuclear recoil energy spectrum was a change of about $\pm15\%$ in bin counts at 1~keVr.  A third source of uncertainty, arising from the Xe(n,n)Xe elastic cross-section data \cite{2009deviveiros}, would appear almost point-like in Fig. \ref{fig1}b and is not shown.  Our $\mathcal{Q}_y$ results were found to be essentially unchanged if the assumed energy resolution was varied by $\pm25\%$, and remained very similar if the fit range was instead truncated at either 10 or 15 electrons.  The agreement between this measurement and that of \cite{2010manzur} (Fig. \ref{fig1}b, $\bigstar$) is quite good above 6~keVr. A possible reason for the rapid rise in $\mathcal{Q}_y$ below 6~keVr obtained in \cite{2010manzur} is discussed in \cite{2010sorensen}.  The previous work \cite{2009sorensen} (Fig. \ref{fig1}b, $\triangledown$) shows systematically higher values of $\mathcal{Q}_y$ below about 25~keVr, possibly due to the systematic effect of {\it false single scatters} mentioned at the beginning of Sec. \ref{qysec}.  It is also worth emphasizing that the previous work \cite{2009sorensen} relied on the S1 signal, and therefore had very limited sensitivity to recoil energies smaller than about 5~keVr.

We have argued that our nuclear recoil data sample does not contain a significant electromagnetic background contamination, and that what little contamination exists can be safely subtracted.  This argument is weakened for events in which no S1 signal was detected, since they have an indeterminate $z$ coordinate for the scatter vertex, and also an indeterminate discrimination parameter (S2/S1).   It is clear from Fig. \ref{fig1}a that most events with no S1 have S2~$\lesssim40$~electrons.   A compelling piece of evidence that such events are in fact elastic nuclear recoils is that their spectral shape is similar to that obtained for elastic nuclear recoils with known S1.  In contrast, the population of events with no S1 in the gamma calibration data shows the expected flat S2 spectral shape in the range S2~$\lesssim40$~electrons.  However, this does not fully exclude the possibility of an additional low-energy background.  The most likely origin of such a background, if it exists, would be low-energy gamma or beta scattering at the liquid surface.  Bulk events would be excluded by virtue of the $\sim$$\mu$m range of $\sim$keV particles in liquid xenon.  

About 95\% of events at the liquid surface are measured to have $0.13\lesssim \sigma_{S2} \lesssim 0.23~\mu$s, as can be seen in Fig. \ref{fig2}a. The distribution of  $\sigma_{S2}$ for bulk events, and events with indeterminate drift time are both roughly Gaussian.  The former shows a $14.7\%\pm2.2\%$ excess of events in the region $0.13\lesssim \sigma_{S2} \lesssim 0.23~\mu$s.  This places an upper limit on the fraction of events with no S1 which might be unaccounted surface background events, rather than bulk elastic nuclear recoils.  The effect of such a background (assumed flat in the range $7-40$ electrons) on our energy calibration would be to increase $\mathcal{Q}_y$ at 8~keVr and 16~keVr by about $0.2$~electrons/keVr, and by half that amount at other energies E$_{nr}<32$~keVr.  This is a very small effect and is not indicated with the other uncertainties in Fig. \ref{fig1}b.

\section{Obtaining the $z$ coordinate of a particle interaction from the S2 width} \label{ediffusion}
\begin{figure}
\centering
\includegraphics[width=0.99\textwidth]{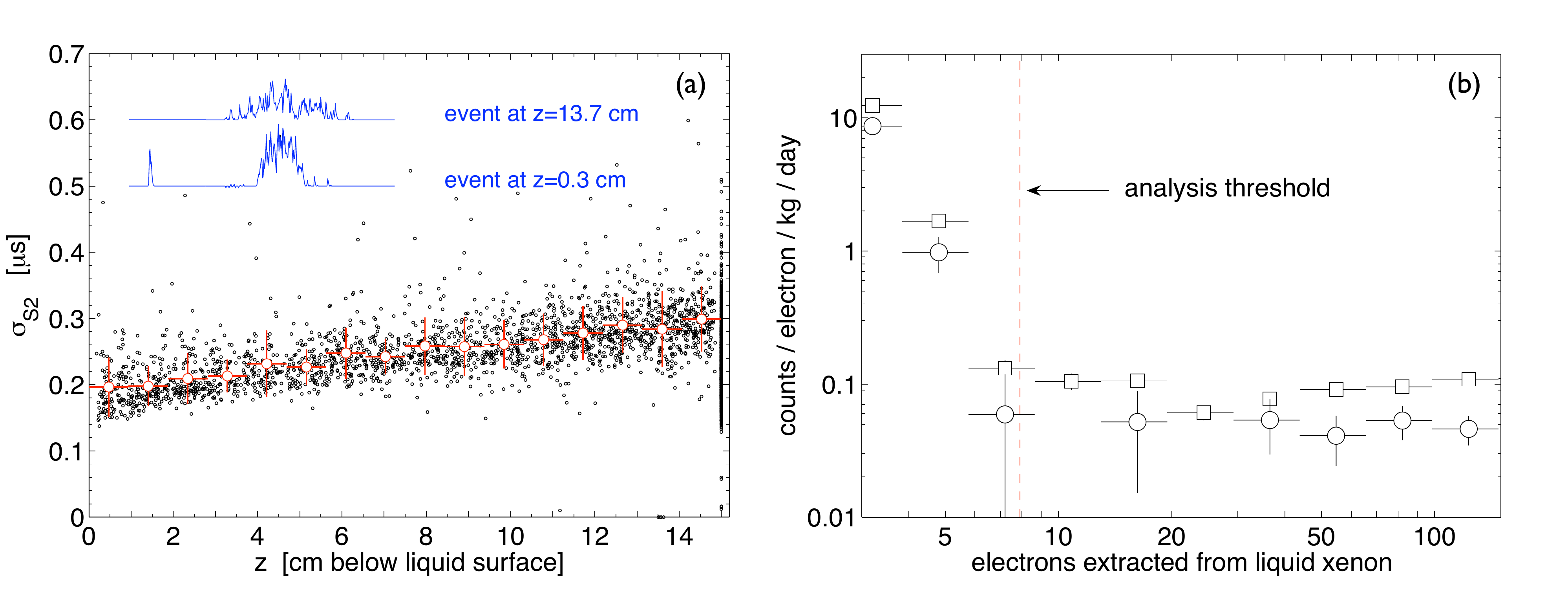}
\caption{The S2 pulse width $\sigma_{S2}$ obtained by Gaussian fit, for single scatter events with $7\leq\mbox{S2}<100$~electrons in the region $r<3$~cm.  The liquid-gas interface is at $z=0$~cm.  Events with indeterminate $z$ coordinate due to absent S1 are shown at z=15~cm.  The band mean is indicated by $\fullmoon$, with $1\sigma$ width;  horizontal bars indicate bin width. A 0.4~$\mu$s section of two typical S2~$\simeq12$~electron events are shown inset (top).  In the event at $z=0.3$~cm, the S1 is visible 1.6~$\mu$s before the S2;  in the event at $z=13.7$~cm, the S1 is far off scale.}
\label{fig2}
\end{figure}

A cloud of ionized electrons resulting from a particle interaction drifts through the liquid xenon target at about $0.20$~cm~$\mu$s$^{-1}$ \cite{1982gushchin}.  As it drifts, its spatial extent broadens due to diffusion.  The amount of diffusion broadening is reflected in the width of the S2 pulse.  In this way, the width of S2 signals depends weakly on the $z$ coordinate of the scatter vertex.  The width $\sigma_{S2}$ was obtained by Gaussian fit to each S2 pulse, and is shown in Fig. \ref{fig2}a for events with $r<3$~cm.  Events with no S1 pulse do not have a precisely defined drift time, and are shown (arbitrarily) at $z=15$~cm.  A 0.4~$\mu$s section of two typical events with S2~$\simeq12$~electrons are shown inset in Fig. \ref{fig2}a, corresponding to nuclear recoils at the extrema (in $z$) of the active xenon target.  The height of the S2 pulse in the $z=0.3$~cm event is about 10~mV referred to the amplifier input.  The $z$ dependence of $\sigma_{S2}$ is clearly weak, however it is sufficient for us to discriminate events which occur at the surface ($z=0$~cm and $z=15$~cm) from those which occur in the bulk of the liquid xenon target.  

\section{Dark matter search data} \label{dmdata}
Figure \ref{fig2}b shows the differential event rate from a 12.5 live day exposure of the XENON10 detector, obtained between August 23 and September 14, 2006.  This data set is distinct from the previously reported \cite{2009angle,2008angle} dark matter search data.  The most notable difference is that during this time, the detector was operated with an S2-sensitive trigger threshold at the level of a single ionized electron.  Event selection criteria were applied as follows.  Valid single scatter events were required to have only a single S2 pulse in the event record.  The efficiency for making this selection is very high, considering the robustness of the S2 signal (as shown in Fig. \ref{fig2}a).  A fiducial cut defined by $r<3$~cm was imposed, giving a target mass of 1.2~kg.  This central region features optimal self-shielding by the surrounding xenon target.  No explicit $z$ cut was made since events were not required to contain an S1 signal.  Events in which an S1 signal was found were required to have log$_{10}$(S2/S1) in the $\pm3\sigma$ band for elastic single scatter nuclear recoils \cite{2009angle}.  Events in which no S1 signal was found were assumed to be low-energy nuclear recoils and were retained.  The fraction of events with no S1 is given by the ratio of the histograms in Fig. \ref{fig1} for the nuclear recoil calibration data. For the dark matter search data, the fraction is approximately $\times3$ lower.  This behavior is expected for electron recoils, based on the observed ratio of log$_{10}$(S2/S1) for electron and nuclear recoils \cite{2009angle}.  The combined acceptance of the two cuts is $\epsilon\gtrsim0.99$, for events in the 1.2~kg central region.

We then used the S2 width $\sigma_{S2}$ to discriminate bulk events from edge events, as described in Sec. \ref{ediffusion}.  In order to ensure complete rejection of events at the liquid xenon surface (at $z=0$~cm), we set the lower bound of the cut at $\sigma_{S2}\geq0.23~\mu$s.  The upper bound of the cut was set at $\sigma_{S2}\leq0.30~\mu$s.  Considering the total ($\sim$Gaussian) distribution of recorded $\sigma_{S2}$ values, these bounds correspond approximately to the region between $\mu$ and $\mu+2\sigma$.  It is clear from Fig. \ref{fig2}a that this only partially targets events from the bottom of the active region of the detector (near $z=15$~cm).  This choice does not compromise surface event rejection, considering that below the cathode grid which defines $z=15$~cm are an additional 1.3~cm of self-shielding liquid xenon.  The acceptance of the S2 width cut for single scatter nuclear recoils is mildly energy-dependent, rising monotonically from $\epsilon=0.39$ for events with S2~=~7~electrons, to $\epsilon=0.44$ for events with S2~$\ge$~40~electrons. 

The S2 spectrum of all single-scatter events that passed these cuts is shown in Fig. \ref{fig2}b ($\fullmoon$).  Vertical bars indicate statistical uncertainty, and horizontal bars indicate bin width.  The count rate is adjusted for the total acceptance fraction $\epsilon$.  The lowest energy events remaining above an analysis threshold of 7 electrons have S2 signals of 8, 17, 18 and 29~electrons.  An additional 4 events were found with $30<\mbox{S2}<44$~electrons. For a higher statistics comparison, the spectrum of all single-scatter events within an 8~cm radius (target mass 8.6~kg) is also shown ($\square$).  No S2 width cut was applied to the data in that case, and the statistical uncertainty is smaller than the data points.  Both spectra are essentially flat above S2~=~7~electrons, as would be expected from a Compton scatter background.  Note that in the calibration analysis of Sec. \ref{qysec}, a lower bound of 7~electrons in the S2 signal was motivated by the trigger efficiency during the neutron calibration.  In the dark matter search data the trigger threshold is at the level of a single electron in the S2 signal.  However, we retain the 7~electron lower bound in analyzing this data because (i) we can most accurately determine acceptance for nuclear recoils above this value, and (ii) the electronic noise increases significantly below S2~$\simeq6$~electrons, as shown in Fig. \ref{fig2}b.  

\section{Summary}
The measured number of S2 electrons can be scaled to nuclear recoil equivalent energy via the $\mathcal{Q}_y$ curve shown in Fig. \ref{fig1}b.  A conservative choice would be the $-1\sigma$ contour, which is approximately $\mathcal{Q}_y\approx4$~electrons/keVr below E$_{nr}=20$~keVr.  As mentioned in Sec. \ref{qysec}, the detector resolution for S2 signals depends primarily on Poisson fluctuations in the number of detected electrons, with an additional component due to instrumental fluctuations.  This is discussed in detail in \cite{2010sorensen}, and for higher energy signals in \cite{2010aprile}.   So as not to overstate the energy resolution, we suggest a parameterization $\mathcal{R}$(keV)~$=1/2\sqrt{\mbox{keV}}$ which follows the Poisson component only.  This is valid for the $-1\sigma$ contour of $\mathcal{Q}_y$.  We have shown that it is possible to give up the usual incident particle type discrimination based on log$_{10}$(S2/S1), and analyze the dark matter sensitivity of XENON10 using only the S2 signal.  The advantage of this analysis appears to be an increased sensitivity to light ($\lesssim10$~GeV) dark matter particles, due to the significantly lower energy threshold.  For larger particle masses, the usual analyses \cite{2009angle, 2008angle} offer superior sensitivity.  Dark matter exclusion limits obtained from the present work should offer substaintial constraints on recent interpretations \cite{2010chang3,2010graham,2010essig,2010hooper,2010fitzpatrick,2010feldstein} of the excess low-energy events observed by CoGeNT \cite{2010aalseth} and CRESST-II \cite{2010seidel}, as well as the DAMA modulation signal \cite{2008bernabei}.

%\section*{Acknowledgments}
%This work benefitted from many discussions during the KITP workshop ``Direct, Indirect and Collider Signals of Dark Matter,'' Santa Barbara CA, December 7-18, 2009, which was supported in part by the National Science Foundation under Grant No. PHY-05-51164.  We gratefully acknowledge support from NSF Grants No. PHY-03-02646 and No. PHY-04-00596, CAREER Grant No. PHY-0542066, DOE Grant No. DE-FG02-91ER40688, NIH Grant No. RR19895, SNF Grant No. 20-118119, FCT Grant No. POCI/FIS/60534/2004 and the Volskwagen Foundation. 

\end{document}